\documentclass{iopconfser}
\usepackage[latin9]{inputenc}
\usepackage{amsmath}
\usepackage{amssymb}
\usepackage{dcolumn}
\usepackage{graphicx}
\usepackage[colorlinks=true, pdfstartview=FitV, linkcolor=blue, citecolor=red, urlcolor=magenta]{hyperref}
\usepackage{amsfonts}
\usepackage{subfigure}
\usepackage{cleveref}
\usepackage{listings}
\usepackage{xcolor}
\usepackage{array}
\usepackage{url}
\usepackage{color}%
\usepackage{tikz}
\usepackage{pgfplots}
\usepackage{todonotes}
\usepackage{multirow}
\usepackage[outdir=./Figures]{epstopdf}

\newcommand{\oo}{\mathring}
\newcommand{\qbar}{\raisebox{-1.5ex}[\height][0pt]{$\mathchar'26$}\mkern-9mu q}

\begin{document}

\title{Near-horizon aspects of black holes in quantum spacetime}

\author{Nikola Herceg$^1$, Tajron Juri\'{c}$^2$, A. Naveena Kumara$^3$, Andjelo Samsarov$^4$ and Ivica Smoli\'{c}$^5$}

\affil{$^{1,2,3,4}$ Rudjer Bo\v{s}kovi\'c Institute, Bijeni\v cka  c.54, 10000 Zagreb, Croatia}
\affil{$^5$ Department of Physics, Faculty of Science, University of Zagreb, 10000 Zagreb, Croatia}

\email{$^1$nherceg@irb.hr, $^2$tjuric@irb.hr, $^3$nathith@irb.hr, $^4$asamsarov@irb.hr, $^5$ismolic@phy.hr}

\begin{abstract}
We give a short introduction to the formalism of noncommutative (twisted) differential geometry that is used to derive the equations of motion for the gravitational perturbation of the Schwarzschild black hole in quantized spacetime. Special attention is given to quantum spacetime arising from $r - \varphi$ noncommutativity.
Tortoise coordinate and near-horizon regions of the effective potentials are analyzed for both polar and axial modes. By carefully examining the associated Schr\"odinger-type equations, we provide the asymptotic solutions at the horizon and illustrate some differences between the polar and axial modes. These findings give further insight into the polar-axial isospectrality violation in the presence of the quantum structure of spacetime.
\end{abstract}

\section{Introduction}
Since the beginning of the 20th century, the unification of quantum theory and gravity has been one of the central problems in theoretical physics.
Given the success of the Standard Model of particle physics, a natural approach was to treat gravity within the framework of quantum field theories that govern the Standard Model.
This was done by quantizing the linearized gravitational field, thereby introducing gravitons -- spin 2 bosons that constitute the gravitational field.

t'Hooft has shown in the 70s \cite{tHooft:1974toh} that divergences arising from the interaction of the gravitational field with a scalar field are non-renormalizable, meaning that infinitely many parameters would be required to regularize them.
The inability to pursue this straightforward route of treating gravity as just another QFT led to the development of alternative approaches to quantum gravity.
Around the '80s, three main approaches to quantizing gravity stood out according to Hawking \cite{Hawking:1979}:\\\\
\textbf{Operator approach}\\
In this approach, the metric in the classical Einstein equation is replaced by a distribution-valued operator on some Hilbert space.
The main obstacles here are the non-polynomiality of the field equations and the difficulty of making sense of the product of the field operators at the same spacetime points, not to mention handling non-polynomial functions such as the inverse metric or the square root of the determinant.\\\\
\textbf{Canonical approach}\\
In this approach, a family of spacelike surfaces foliates the spacetime, allowing the introduction of Hamiltonian and canonical equal-time commutation relations.
The main advantages of this approach are its applicability to strong gravitational fields and manifest unitarity of the theory.
However, one disadvantage is that the topology of the spacetime should be a product of the real line and a spatial part, which limits the richness of the solution set of the field equations. Additionally, the meaning of the equal-time commutation relations is unclear -- these are well defined for matter fields on a fixed spacetime geometry, but what sense does it make to say that two points are spacelike separated if the geometry is quantized and obeys the uncertainty principle?\\\\
\textbf{Path integrals}\\
The starting point in this approach is Feynman's idea that one can represent the amplitude to go from a state with a metric $g_1$ and matter fields $\phi_1$ on a surface $S_1$ to a state with a metric $g_2$ and matter fields $\phi_2$ on a surface $S_2$ as a sum over all field configurations $g$ and $\phi$ which take the given boundary values at the surfaces $S_1$ and $S_2$:
\begin{equation*}
\langle g_2,\phi_2, S_2 | g_1, \phi_1, S_1 \rangle = \int \mathcal D [g, \phi] \exp (i S[g, \phi]).
\end{equation*}

This approach, pioneered by Hawking, had wide-ranging success in areas ranging from cosmology to black hole thermodynamics.
\\\\

While some of the above approaches have been more successful than others, none of them managed to merge quantum theory and gravity in a fully satisfactory manner. The quest for quantum gravity continued and not long after, string theory \cite{Green:2012oqa} and loop quantum gravity \cite{Rovelli:2004tv} emerged as promising background-independent theories of quantum gravity.

Noncommutative (NC) geometry is another approach to quantum gravity that became increasingly popular in the 90s. The mathematics behind it came from two important directions -- quantum group theory and spectral geometry a la Connes \cite{Connes:1995tu}.
From a physicist's perspective, noncommutative geometry presented another venue for spacetime quantization.

Noncommutative spaces can be realized as certain limits of string theory and loop quantum gravity models.
By making the algebra of functions on a spacetime manifold noncommutative, one introduces fuzziness at the coordinate level.
In the next section, we will present an approach to NC geometry that is based on a Hopf algebra and star-product of functions on a spacetime manifold.

If we were to fit this approach into one of the three categories mentioned above, it would be the first one -- the operator approach. While this might seem unintuitive at first glance, the connection between star-products and Hilbert space formalism can be made explicit by utilizing the Wigner-Weyl transforms. These transformations form the foundations of the phase space formulation of quantum mechanics and serve as a tool for transforming the Hilbert space operators and density matrices into the ordinary phase space functions and vice versa \cite{Szabo:2001kg}.

The idea behind phase space quantization in quantum mechanics is to encode the noncommutative operator structure of the position and momentum observables (and their functions, such as Hamiltonian) into the star-product of ordinary functions on the phase space.
In our setting, it is not the phase space that exhibits noncommutativity, but rather the spacetime itself, thereby replacing the aforementioned phase space star-product with the coordinate space star-product.

The theory of noncommutative gravity presented in the following section partially mitigates the issues mentioned under the first point in the approaches above. The main simplification is that the operator structure is solely in the product of functions; that is, $\sqrt{\text{det }g}$ is the same object in both the quantized and unquantized theories. The field equation is obtained through a bottom-up approach, avoiding complexities arising from the action and variational calculus. Everything is done up to linear order in gravitational perturbation and noncommutative parameter.

After introducing the formalism of NC differential geometry, we apply it to the case of linearized black hole perturbation theory. Perturbations of the black hole metric are, as in the classical case, governed by a single Schr\"odinger-type equation. Effects of noncommutativity are encoded in corrections to the classical potentials and tortoise coordinates. 
We will focus on the $\qbar$-space \cite{Herceg:2023zlk} for which $[\varphi \stackrel{\star}{,}r]=i\qbar$ in spherical coordinates.
Special attention is given to the near-horizon region, where NC effects lead to \textit{mode-dependent} translation of the horizon and requirement of regularity gives us the precise location of the horizon up to the first order in $\qbar$. Differences between the polar and axial NC potentials in the near-horizon region shed some light on spectral discrepancies of the potentials, giving further insight into polar-axial isospectrality violation calculated in \cite{Herceg:2024vwc}.

\section{Noncommutative differential geometry}

In this section, we will provide a brief introduction to Hopf algebras and twisted differential geometry based on \cite{Aschieri:2005zs, Aschieri:2005yw, Aschieri:2009qh, schenkel}.
Hopf algebra arises as a central algebraic structure of many approaches to noncommutative physics.
Its role is similar to that of a gauge group in gauge theory -- it encodes the symmetries of the system.
While gauge group deals with internal symmetries of the fields, Hopf algebra modifies the symmetries of the spacetime manifold itself at the coordinate level. 

Symmetries of a noncommutative spacetime are governed by the Hopf algebra of deformed diffeomorphisms $H^\mathcal{F}$. It is constructed by twisting the coalgebraic sector of the Hopf algebra of diffeomorphisms $H$ using a twist element (2-cocycle) $\mathcal F$. 
To understand the structure of $H$ we start by considering a Lie algebra of diffeomorphisms $(\Xi, [\cdot, \cdot])$.
$\Xi$ can be embedded into a larger structure, the universal enveloping algebra $(U \Xi, \mu, \eta)$, which is endowed with associative product $\mu : U \Xi \otimes U \Xi \to U \Xi$ and a unit map $\eta : \mathbb{C} \to U \Xi$. The Hopf algebra $H$ is then formed by further adding the coproduct $\Delta : H \to H \otimes H$, counit $\epsilon : H \to \mathbb{C}$ and antipode $S : H \to H$. 

Compatibility of these structures is required, e.g. compatibility of the product and coproduct is given by a bialgebra property of the Hopf algebra \cite{Majid:1996kd}.
On a vector space level, the constructed Hopf algebra $H$ is isomorphic to the enveloping algebra $U \Xi$.
The co-structures materialize the concepts of Leibniz rule (coproduct), normalization (counit) and inverse (antipode) at the diffeomorphism level, without any reference to the algebra of functions $ \mathcal A = (C^\infty (\mathcal{M}), \cdot)$ that the Hopf algebra acts on. Coproduct in $H$ is given as $\Delta h = h \otimes 1 + 1 \otimes h, \ \forall h \neq 1$.
The Leibniz rule can be written as 
\begin{equation} \label{leibniz}
h \rhd (f \cdot g) =  \cdot \ \Delta h (\rhd \otimes \rhd) (f \otimes g),
\end{equation}
where $\rhd$ is a Lie derivative action of $H$ on functions in $\mathcal A$.

Twisting of $H$ into $H^\mathcal{F}$ is performed using a Drinfeld twist element $\mathcal F \in H \otimes H$ by modifying the co-structures, thus going from $H = (U \Xi, \mu, \eta, \Delta, \epsilon, S)$ to $H^\mathcal{F} = (U \Xi, \mu, \eta, \Delta^\mathcal{F}, \epsilon, S^\mathcal{F})$. Specifically, the coproduct gets twisted as $\Delta^\mathcal{F} h = \mathcal{F} \Delta h \mathcal{F}^{-1}$.

Since the coproduct appears on the RHS of the Leibniz rule \eqref{leibniz}, twisting the coproduct implies that we have to twist the product $\cdot$ in $\mathcal A$ as well in order to keep the Leibniz rule. Therefore we deform the pointwise product $\cdot$ into the star-product $\star$ defined as
\begin{equation} \label{starproduct}
f \star g :=\cdot \  \mathcal{F}^{-1}(\rhd \otimes \rhd) (f \otimes g).
\end{equation}
The Leibniz rule now holds as can be explicitly checked by expanding $h \rhd (f \star g) = \star \ \Delta^\mathcal{F}h \ (\rhd \otimes \rhd) (f \otimes g)$. Later on, we will work in spherical coordinates, so we use the twist of the form
\begin{equation} \label{semipseudo}
\mathcal{F} = \exp \big(-i\frac{\qbar}{2} ( \partial_\varphi \otimes \partial_r - \partial_r \otimes \partial_\varphi) \big).
\end{equation}
In order to evaluate this exponential, the field $\mathbb{C}$ must be upgraded to a formal power series in NC parameter $\mathbb{C}[[\qbar]]$. We can therefore use the notation $\mathcal{F} = f^\alpha \otimes f_\alpha, \ \mathcal{F}^{-1} = \overline f^\alpha \otimes \overline f_\alpha$, where sum over repeated indices is implied. 
The star-product up to the first order in $\qbar$ is
\begin{equation} \label{starpr}
f \star g = \overline f^\alpha(f) \overline f_\alpha (g) = f \cdot g + \frac{i \qbar}{2} \big(  \partial_\varphi f \partial_r g -  \partial_\varphi g \partial_r f \big) + \mathcal O (\qbar^2),
\end{equation}
where we used the notation $\overline f^\alpha(f) := \overline f^\alpha \rhd f$.
The only non-zero commutator between the functions in the spherical coordinate chart is $[\varphi \stackrel{\star}{,}r] = i \qbar$.
Algebra of functions $\mathcal A = (C^\infty(\mathcal M), \cdot)$ thus deforms into $\mathcal A_\star = (C^\infty(\mathcal M), \star)$, where $\qbar$ plays the role of NC parameter.

The Lie derivative, tensor product and pairing of vector fields and 1-forms can be deformed consistently by following a general prescription of precomposing the undeformed operation of interest with the inverse twist action, as in the star-product \eqref{starproduct}. 
This process ensures the property of star-linearity, i.e. tensors are $\mathcal A_\star$-modules.
We therefore have
\begin{align*}
\tau \otimes_\star \tau' :=&\ \overline f^\alpha (\tau) \otimes \overline f_\alpha (\tau '), \\
\pounds^\star_v f :=&\ \pounds_{\overline f^\alpha (v)} (\overline f_\alpha(f)), \\
\langle \omega, v \rangle_\star :=&\ \langle \overline f^\alpha (\omega), \overline f_\alpha (v) \rangle,
\end{align*}
where $\tau, \tau'$ are general tensor fields, $f$ is a function, $v$ is a vector field and $\omega$ is a 1-form. It is important to note that the sets of all functions and tensor fields remain the same upon deformation -- it is just the algebraic structure that gets deformed.

The deformed covariant derivative is a star-linear mapping $\hat \nabla_v : \Xi \to \Xi$ along the vector field $v$ that satisfies the following axioms:
\begin{equation} \label{covariantder}
\begin{aligned}
       \hat{\nabla}  _{v+w} \: z=& \: \hat{\nabla}  _v z+\hat{\nabla}  _w z ,\\
\hat{\nabla}  _{h\star v} \: z=& \: h \star \hat{\nabla}  _v z, \\
        \hat{\nabla}  _{v}(h \star z)=& \: \pounds_v^\star (h) \star z+\bar R ^\alpha (h)\star \hat{\nabla}  _{\bar R_\alpha (v)}z,
\end{aligned}  
\end{equation}
where $\mathcal R = \overline R^\alpha \otimes \overline R_\alpha$ is the universal $\mathcal R$-matrix that satisfies the quantum Yang-Baxter relations \cite{Majid:1996kd}.
The appearance of the $\mathcal R$-matrix in the last property reflects the braiding structure that underlies the whole theory. The $\mathcal R$-matrix can be obtained from the twist $\mathcal{F}$ as $\mathcal R = \mathcal{F}_{\text{op}} \mathcal{F}^{-1} = f_\alpha \overline f^\beta \otimes f^\alpha \overline f_\beta$. The star-product is actually $\mathcal R$-symmetric, i.e. $f \star g = \overline R^\alpha(g) \star \overline R_\alpha(f)$. This action of the $\mathcal R$-matrix upon swapping is visible in the third identity in \eqref{covariantder}, where the function $h$ and vector $v$ on the RHS swap places relative to their ordering on the LHS, hence the action of the $\mathcal R$-matrix on the RHS.

The deformed torsion and Riemann tensors are defined by 
\begin{align}  \label{torsion-riemann}
  \hat{T}(u,v) &:=  \hat{\nabla}_{u} v - \hat{\nabla}_{\bar{R}^{c} (v)} {\bar{R}_{c} (u)}
  -  [u, v]_{\star}   \equiv {\langle  u \otimes_{\star} v,  \hat{T}   \rangle}_{\star},  \\
 \hat{R}(u,v,z) &:=  \hat{\nabla}_{u}\hat{\nabla}_{v} z
  -  \hat{\nabla}_{\bar{R}^{c} (v)}\hat{\nabla}_{\bar{R}_{c} (u)} z -
  \hat{\nabla}_{[u, v]_{\star}} z   \equiv {\langle  u \otimes_{\star} v \otimes_{\star} z,  \hat{R}   \rangle}_{\star},
\end{align}
where $[u ,v]_\star := \pounds^\star_u (v)$. Both of these tensors are $\mathcal R$-antisymmetric in the first two slots, e.g. $\hat T(u, v) = - \hat T(\overline R^\alpha (v), \overline R_\alpha (u))$.
The Ricci tensor is given by
\begin{equation}
\hat{R}(u,v) := \langle dx^\alpha, \hat{R}(\partial_\alpha, u, v) \rangle_\star.
\end{equation}
This deformed Ricci tensor does not exhibit the $\mathcal R$-symmetry that we would expect based on its classical counterpart. The \textit{naive} vacuum Einstein manifold condition $\hat R_{\mu \nu} = 0$ would lead to an overcomplete system of equations that imply $\qbar = 0$. This inspired the proposal \cite{Herceg:2023zlk,Herceg:2023pmc}
\begin{equation} \label{einstein}
\hat{{\rm R}}_{\mu\nu} := \frac{1}{2}\left\langle dx^{\alpha}, \hat{R}(\partial_{\alpha}, \partial_{\mu}, \partial_{\nu})+\hat{R}(\partial_{\alpha}, \bar{R}^{A}(\partial_\nu), \bar{R}_{A}(\partial_\mu)\right\rangle_\star = 0
\end{equation}
for the vacuum Einstein manifold that allows for nontrivial contributions coming from the spacetime noncommutativity, as we will see in the next section.

\section{Scattering on a noncommutative black hole}
In this section, we apply the formalism presented in the previous section within the context of gravitational perturbation theory.
We examine the linear perturbations of the Schwarzschild metric on a noncommutative background.
Two main ingredients are the background metric $\oo g$ and the twist $\mathcal{F}$.
The background metric $\oo g$ is Schwarzschild,
\begin{equation}
	d s^2=-\left(1-\frac{R}{r}\right) c^2 d t^2+\frac{1}{1-R / r} d r^2+r^2\left(d \theta^2+\sin ^2 \theta d \varphi^2\right), \qquad
	R = \frac{2 G M}{c^2},
\end{equation}
and the twist $\mathcal F$ is given in \eqref{semipseudo}. One technical advantage of this twist is the fact that the vector fields constituting it act trivially on all elements of the spherical coordinate vector field basis $(\partial_t, \partial_r, \partial_\theta, \partial_\varphi)$.
This makes the calculations much simpler when using the spherical basis for the tensor components, e.g. the inverse metric property is simply
\begin{equation*}
	g^{\mu \nu}_\star \star g_{\nu \rho} = \delta^\mu_{~ \rho}, \quad g_{\mu \nu} \star g^{\nu \rho}_\star = \delta_{\mu}^{~ \rho},
\end{equation*}
and metric compatible Levi-Civita connection is given by 
 \begin{equation}\label{gama} 
  \hat{\Gamma}^{\mu}_{~\nu \rho} = \frac{1}{2} g^{ \mu \alpha}_\star \star \big( \partial_{\nu} g_{\rho \alpha}  +  \partial_{\rho} g_{\nu \alpha} -  \partial_{\alpha} g_{\nu \rho} \big),  
\end{equation}
where the connection coefficients come from action on the basis vectors, 
\begin{equation} \label{nablamu}
\hat{\nabla}_{\partial_\mu} \partial_{\nu} = \hat{\Gamma}^{\lambda}_{~\mu \nu} \star \partial_{\lambda} = \hat{\Gamma}^{\lambda}_{~\mu \nu} \partial_{\lambda}.
\end{equation}
The last equality in \eqref{nablamu} follows from triviality of the action of the twist $\mathcal F$ on the basis vectors.

To find the noncommutative corrections to the black hole perturbations, we split the metric $g = \oo g + h$ into the background part and perturbation. Then we proceed with finding the metric inverse, connection, Riemann tensor and the Ricci tensor, while keeping only terms linear in perturbation and NC parameter $\qbar$. Finally, we impose \eqref{einstein} to obtain the equations of motion.

Due to the combination of radial derivative and Killing field of the background metric $\oo g$ inside the twist \eqref{semipseudo}, the noncommutative corrections appear exclusively together with the perturbation $h$. This is easily seen by evaluating generic star-products $\oo g \star h$ and $ \oo g \star \oo g$ while using the fact that $\pounds_{\partial_\varphi}(\oo g) = 0$.

A specific ansatz for $h$ must be used in order to solve the vacuum NC Einstein equation. Given the spherical symmetry of the background, the perturbation naturally decomposes into tensor spherical harmonics as shown by Regge and Wheeler. Two classes of perturbation arise, characterized by their behavior with respect to the parity operator $\vec{r} \mapsto -\vec{r}$.

\begin{figure}[!h]
	\subfigure[]{\includegraphics[scale=0.50]{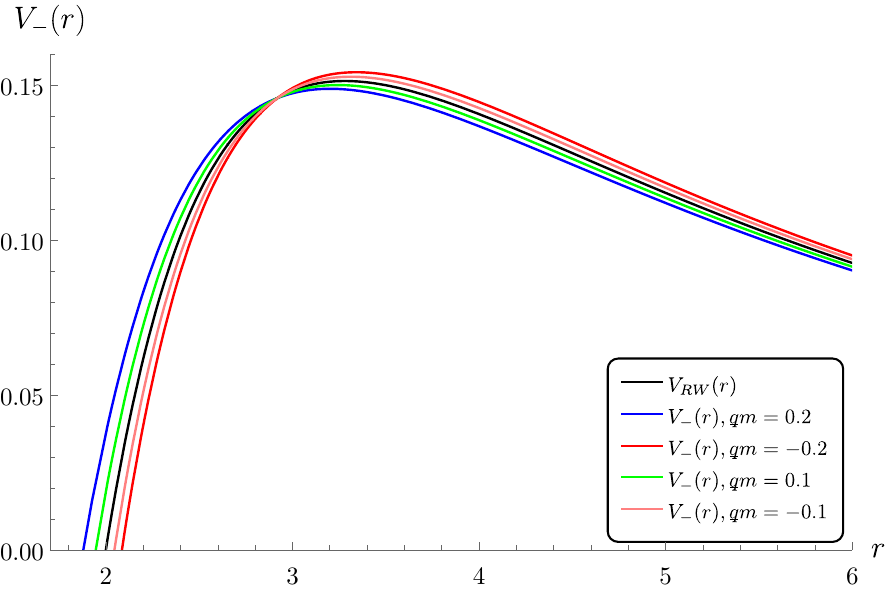}\label{figRW}} \qquad
	\subfigure[]{\includegraphics[scale=0.50]{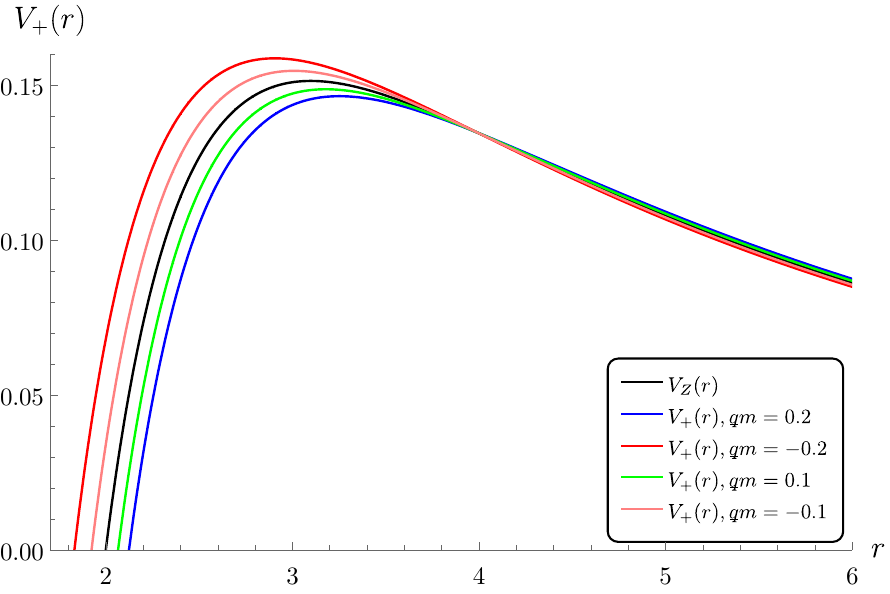}\label{figZ}}  \\
	\subfigure[]{\includegraphics[scale=0.50]{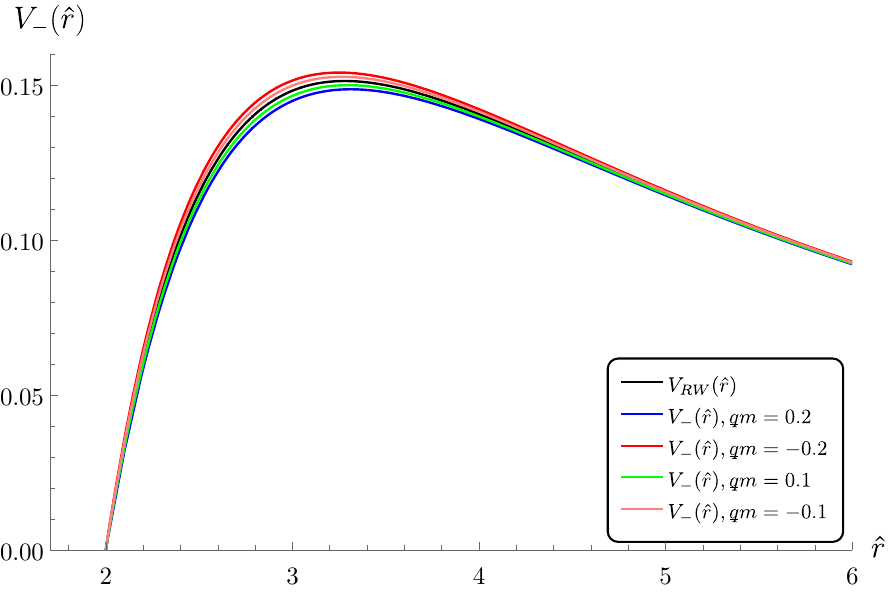}\label{figRWhat}} \qquad
	\subfigure[]{\includegraphics[scale=0.50]{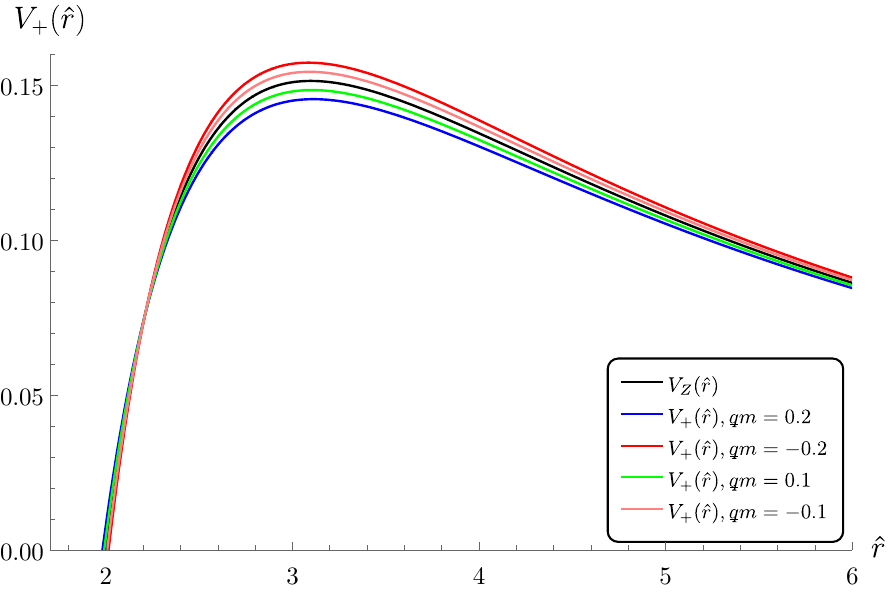}\label{figZhat}}  \\
	\subfigure[]{\includegraphics[scale=0.50]{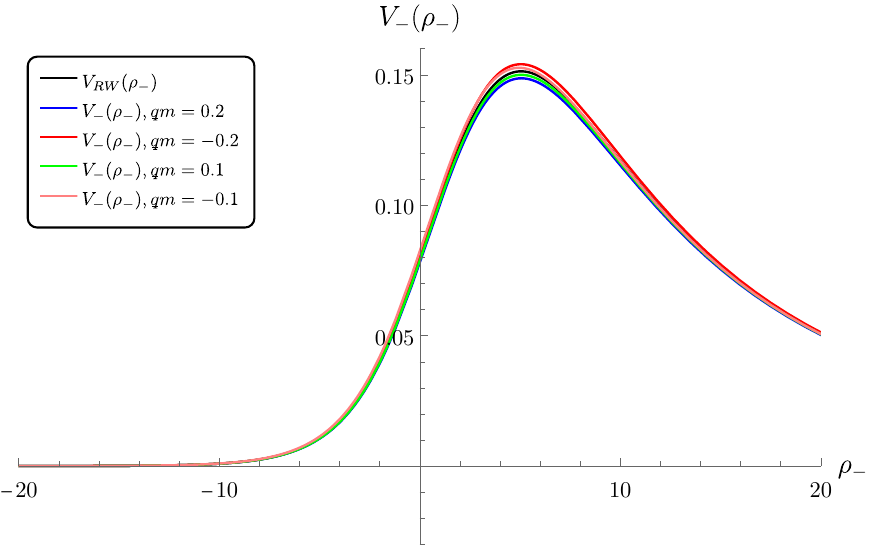}\label{figRWtort}} \qquad
	\subfigure[]{\includegraphics[scale=0.50]{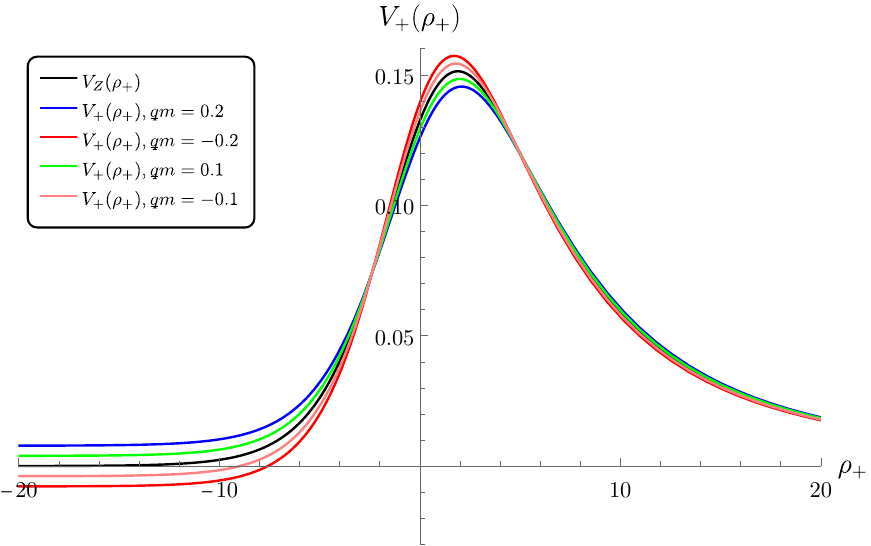}\label{figZtort}}
	\caption{Noncommutative polar and axial potentials in $r, \hat r$ and $\rho$ coordinates for $\ell = 2$. Axial potential is in the left column and polar potential is in the right.}
\end{figure}

\subsection{Axial modes} 
Axial modes are those with a negative eigenvalue with respect to the parity operator. 
The effective potential governing the axial modes was derived by Regge and Wheeler \cite{rw} and was later generalized to the case of a noncommutative black hole in \cite{Herceg:2023zlk,Herceg:2023pmc} where the details of the results presented in this subsection can be found. Specifically, the $\qbar$-space analyzed here has been studied in \cite{Herceg:2023zlk}.
The perturbation in the Regge-Wheeler gauge is decomposed as
\begin{equation*}
\begin{split}
	h_{t \theta}&=\frac{1}{\sin \theta} \sum_{\ell, m} h_0^{\ell m} \partial_{\varphi} Y_{\ell m}(\theta, \varphi)e^{-i \omega t}, \qquad
	 h_{t \varphi}=-\sin \theta \sum_{\ell, m} h_0^{\ell m} \partial_\theta Y_{\ell m}(\theta, \varphi)e^{-i \omega t},\\
	h_{r \theta}&=\frac{1}{\sin \theta} \sum_{\ell, m} h_1^{\ell m} \partial_{\varphi} Y_{\ell m}(\theta, \varphi) e^{-i \omega t}, \qquad
	 h_{r \varphi}=-\sin \theta \sum_{\ell, m} h_1^{\ell m} \partial_\theta Y_{\ell m}(\theta, \varphi) e^{-i \omega t}.
\end{split}
\end{equation*}

Imposing the NC Einstein equation \eqref{einstein} produces three distinct coupled ordinary differential equations in $r$ after separating out the angular part.
Solving the system amounts to reducing it to a single Schr\"odinger-type equation as presented in \cite{Herceg:2023zlk,Herceg:2023pmc}:
\begin{equation}
	\frac{d^2 \Psi}{d \rho_-^2} + \big(\omega^2 - V_-(r) \big)\Psi = 0,
\end{equation}
where
\begin{align}
	\rho_- &= r + R \log \frac{r - R}{R} + \frac{\qbar m}{2} \frac{R}{r - R}, \label{RWtortoise} \\
	V_-(r) &= \frac{(r - R)\big(\ell (\ell + 1)r - 3R\big)}{r^4} + \qbar m \frac{\ell(\ell + 1)(3R - 2r)r + R(5r - 8R)}{2 r^5}, \label{ncRW}
\end{align}
are the axial NC tortoise coordinate and NC Regge-Wheeler potential, respectively. 

There are two competing terms in \eqref{RWtortoise}. Classically, the logarithm term sends the tortoise coordinate to $-\infty$ as we approach the horizon at $r = R$. 
However, since $1/x$ diverges more rapidly than $\log x$ near $x = 0$, the NC correction dominates the coordinate transformation near the horizon. Furthermore, the transformation becomes ill-defined for positive values of $\qbar m$ near the horizon since there is a point where the Jacobian of the transformation \eqref{RWtortoise} vanishes.

The potential \eqref{ncRW} is on the other hand regular at the horizon. Classically, the potential is zero at the horizon as can be seen from the commutative part. When we include the NC correction, the zero of the potential is translated to $r = \hat R := R - \qbar m / 2 + \mathcal O (\qbar^2)$. It is therefore suggestive to use the shifted radial coordinate $\hat r = r + \qbar m / 2 $ that satisfies $r = \hat R$ for $\hat r = R$.

We can now reexpress the tortoise coordinate and the potential using $\hat r$,
\begin{align}
	\rho_- &= \hat r + R \log \frac{\hat r - R}{R} - \qbar m / 2 , \label{tortoise_axial_hat} \\
	V_-(\hat r) &= \frac{(\hat r - R)\big(\ell (\ell + 1)\hat r - 3R\big)}{{\hat r}^4} - \qbar m \frac{(\hat r-R)R}{{\hat r}^5}. \label{ncRWhat}
\end{align}

Now, the tortoise coordinate exhibits standard behavior and the potential vanishes at $\hat r = R$ for all $\qbar m$. The plot of the potential in terms of $r, 
\hat r$ and $\rho_-$ is given in Figures \ref{figRW}, \ref{figRWhat} and \ref{figRWtort} respectively.
In conclusion, for the axial case, the coordinate $\hat r = r + \qbar m / 2$ accomplishes two things: it regularizes the tortoise coordinate transformation at the translated horizon ($\rho_- \to -\infty$ for $\hat r \to r$), and the potential is zero at the translated horizon as in the classical case. In the polar case, we will see that there exists no single coordinate $\hat r$ that satisfies both properties.

\subsection{Polar modes}
The polar modes are even with respect to the parity operator $r \to -r$. Effective potential governing the polar modes has been found by Zerilli \cite{zer} and was generalized to the noncomutative setting in \cite{Herceg:2024vwc}. Here, we provide a brief overview of the derivation. Details of the derivation and elaborate analysis of the QNM spectrum can be found in \cite{Herceg:2024vwc}.
The perturbation $h$ for this set of modes can be written in Zerilli gauge as
\begin{eqnarray*}
\label{eq:even-pert2}
&&h_{tt} = f(r)\sum_{\ell, m} H_{0}^{\ell m}(r) Y_{\ell m}(\theta,\varphi)e^{-i \omega t},  \quad
h_{tr} = \sum_{\ell, m} H_{1}^{\ell m}(r) Y_{\ell m}(\theta,\varphi)e^{-i \omega t}, \\
&&h_{rr} = \frac{1}{f(r)} \sum_{\ell, m} H_{2}^{\ell m}(r) Y_{\ell m}(\theta,\varphi)e^{-i \omega t},  \quad 
h_{ab} = \sum_{\ell, m} K^{\ell m}(r) \oo g_{ab} Y_{\ell m}(\theta,\varphi)e^{-i \omega t}.
\end{eqnarray*}

Imposing the NC Einstein equation \eqref{einstein} produces 7 distinct differential equations in $r$. The system can be solved by reducing it to Schr\"ordinger-type equation, which again requires introduction of the tortoise coordinate 
\begin{equation} \label{Ztortoise}
	\rho_+ =\ r + R \log \frac{r - R}{R} - \qbar m \left(\frac{(2 \Lambda +7) R}{2 (2 \Lambda +3)
   (r-R)}-\frac{4 (\Lambda +3) \log
   \left(r/R-1\right)}{(2 \Lambda +3)^2}-\frac{9 \log
   \left(2 \Lambda r/R+3\right)}{\Lambda  (2 \Lambda
	+3)^2}\right),
\end{equation}
where $2 \Lambda = \ell(\ell + 1)-2$. The potential is 
    \begin{align}
        V_+=&\ \frac{(r-R) \left(8 \Lambda ^2 (\Lambda +1) r^3+12 \Lambda ^2 r^2 R+18 \Lambda  r R^2+9 R^3\right)}{r^4 (2 \Lambda  r+3 R)^2} \nonumber \\
	    & + \frac{\qbar m}{4 r^5 (2 \Lambda  r+3 R)^3} \Big(32 \Lambda ^2 \left(2 \Lambda ^2+7\right) r^5-8 \Lambda ^2 (2 \Lambda  (6 \Lambda -13)+121) r^4 R  \\
	    & -12 \Lambda  (2 \Lambda  (15 \Lambda -58)+59) r^3 R^2-2 (\Lambda  (440 \Lambda -741)+162) r^2 R^3 -3 (316 \Lambda -207) r R^4-387 R^5 \Big).\nonumber
    \end{align}
Tortoise coordinate transformation again displays similar pathology when $r \sim R$. The first term in the brackets of \eqref{Ztortoise} competes with $\log(r/R-1)$ and causes problems for negative values of $\qbar m$. This can again be addressed by introducing $\hat r = r - \qbar m \frac{2 \Lambda + 7}{4 \Lambda + 6}$. Tortoise coordinate and potential in terms of $\hat r$ are
\begin{equation*}
	\begin{aligned}
		&\rho_+ = {\hat r} + R \log  \frac{{\hat r}-R}{R}+ \qbar m \left(\frac{2 \Lambda +7}{4 \Lambda +6}+\frac{4 (\Lambda +3) \log
   \left(\hat r / R-1\right)}{(2 \Lambda +3)^2}+\frac{9 \log \left(2 \Lambda  {\hat r}/R+3\right)}{\Lambda  (2 \Lambda +3)^2}\right), \\
		&V_+(\hat r) = \frac{({\hat r}-R) \left(8 \Lambda ^2 (\Lambda +1) {\hat r}^3+12 \Lambda ^2 {\hat r}^2 R+18 \Lambda  {\hat r} R^2+9 R^3\right)}{{\hat r}^4 (2 \Lambda  {\hat r}+3 R)^2} \\
                &+\frac{\qbar m}{4 (2 \Lambda +3) {\hat r}^5 (2 \Lambda  {\hat r}+3 R)^3}
                \Big(
        96 \Lambda ^2 \left(7-4 \Lambda ^2\right) {\hat r}^5+8 \Lambda ^2 \left(88 \Lambda ^2-206 \Lambda -363\right) {\hat r}^4 R+   6 \Lambda  (2 \Lambda
                (\Lambda  (10 \Lambda +159)-67) \\ &-177) {\hat r}^3 R^2 +2 (\Lambda  (2 (765-188 \Lambda ) \Lambda +513)-486) {\hat r}^2 R^3+3 \left(-272 \Lambda ^2+618
   \Lambda +243\right) {\hat r} R^4+9 (39-38 \Lambda ) R^5
                \Big).
		\end{aligned}
\end{equation*}

Noncommutative Zerilli potential is plotted in Figures \ref{figZ}, \ref{figZhat}, \ref{figZtort} in terms of $r, \hat r, \rho$ respectively. 
Unlike the axial case, here there is no unique mode-dependent coordinate $\hat r$ that makes the potential zero at the horizon and regularizes the tortoise coordinate at the same time. The choice of $\hat r = r - \qbar m \frac{2 \Lambda + 7}{4 \Lambda + 6}$ achieves the usual behavior of the tortoise coordinate, i.e. $\rho_+ \to -\infty$ monotonically as $\hat r \to R$. The asymptotic value of the NC potential at $\rho_+ \to -\infty$ is 
\begin{equation}
	\lim_{\rho_+ \to -\infty} V_+(\rho_+)=\qbar m \frac{ 55\Lambda ^3-144 \Lambda ^2+117 \Lambda +9}{(2 \Lambda +3)^3 R^3},
\end{equation}
which can also be seen from Fig. \ref{figZtort}. This qualitative difference relative to the axial case points to violation of the classical polar-axial isospectrality.

\section{Outlook and discussion}
After a brief introduction to NC differential geometry arising from the Hopf algebra of deformed diffeomorphisms, we applied the theory to the case of polar and axial gravitational perturbations of the Schwarzschild black hole. As in the classical case, solutions are governed by Schr\"odinger-type equations with specific effective potentials and tortoise coordinates.

Quasinormal modes are solutions of these equations that are subject to the boundary condition of purely ingoing waves at the horizon, that is $\Psi_\pm \propto e^{-i k \rho_\pm}$ as $\rho_\pm \to -\infty$ and purely outgoing at the spatial infinity, $\Psi_\pm \propto e^{i k \rho_\pm}$ for $\rho_\pm \to \infty$. 

In the axial case, the wavenumber $k$ is equal to $\omega$ as in the classical case since the potential vanishes in both asymptotic regions. In the polar case, however, potential tends to a nonzero value at the horizon and therefore the solution is 
\begin{equation}
	\Psi_+ \propto e^{- i k \rho_+}, \quad k^2 = \omega^2 + \qbar m \frac{ 55\Lambda ^3-144 \Lambda ^2+117 \Lambda +9}{(2 \Lambda +3)^3 R^3} \text{ for } \rho_+ \to -\infty.
\end{equation}
This can be interpreted as a modified dispersion relation for the polar perturbations in the near-horizon region, further illustrating the violation of isospectrality between the axial and polar modes. In \cite{Herceg:2024vwc} it has been calculated that for polar modes (+), negative values of $\qbar m$ lead to an increase in negative imaginary part of the QNM frequency (stronger damping) compared to the axial (and commutative) frequency and vice versa for negative $\qbar m$ values. This behavior is expected if we look at Fig. \ref{figZtort}. The red line on that figure has a higher peak and negative asymptotic value of the potential. From the wave's perspective, a higher peak corresponds to a higher potential barrier and negative asymptotic potential implies stronger attraction by the black hole, both phenomena leading to increased damping. The opposite logic holds for the positive $\qbar m$ values (blue line).

In the end, it is worth pointing out that the precise location of the horizon is obtained by identifying the pole in the tortoise coordinate transformation. Direct transformation from $r$ to $\rho_\pm$ is invalid at the first order and intermediate radial coordinate $\hat r$ has been introduced to absorb this divergence into the single logarithm that naturally stretches the exterior region to $-\infty$ for all values of $\qbar m$. 

\section*{Acknowledgment}
This  research was supported by the Croatian Science Foundation Project No. IP-2020-02-9614 \textit{Search for Quantum spacetime in Black Hole QNM spectrum and Gamma Ray Bursts}. Part of the calculations were checked using the Mathematica package.

\end{document}